\documentclass{article}
\usepackage{graphicx}
\begin{document}
\title{Propagation of light in low pressure ionised and atomic hydrogen. Application to astrophysics.
\author{Jacques Moret-Bailly 
\footnote{Laboratoire de physique, Universit\'e de Bourgogne, BP 47870, F-21078 Dijon cedex, France. 
email : Jacques.Moret-Bailly@u-bourgogne.fr}}}
\maketitle

\begin{abstract}
 \textquotedblleft Impulsive Stimulated Raman Scattering\textquotedblright (ISRS) uses ultra-short laser 
pulses to shift light frequencies; the frequency shift depends on the power of the laser pulses because this 
power is very large. 
The relative frequency shifts of \textquotedblleft Coherent Raman Effect on Incoherent 
Light\textquotedblright (CREIL) described in this paper are independent on the intensity of the ordinary 
incoherent light that it uses, and, in a first approximation, on the frequency of the light. Since CREIL does 
not blur images or alter the spectral pattern, CREIL effect may be confused with Doppler frequency shifts. 
ISRS and CREIL are parametric effects that do not excite matter, they transfer energy from 
\textquotedblleft hot beams\textquotedblright to \textquotedblleft cold beams\textquotedblright . These 
transfers correspond to spectral shifts; in CREIL thermal radiation is blue-shifted, that is heated.

CREIL requires low pressure gases acting as catalysts. These gases must have Raman transitions in the 
radio frequencies range: for example H$_2^+$ or excited atomic hydrogen in a magnetic field.

The spectral lines resulting from a simultaneous absorption (or emission) and CREIL have a width at least 
equal to the frequency shift, so that the lines of a complex spectrum may be weakened and mixed, 
becoming nearly invisible.

In interstellar space, molecular hydrogen is ionised, but since H$_2^+$ is quickly destroyed by collisions it 
persists only at pressures low enough to provide CREIL; the redshift widens the weak absorption lines of 
H$_2^+$ which becomes undetectable. It contributes to the \textquotedblleft cosmological 
redshift\textquotedblright and amplification of the microwave 2.7K background radiation.

Using only well established physics and normal astronomical objects, CREIL provides a plausible 
explanation for the enigmatic spectra of the quasars. 

\end{abstract}
keywords: plasma, quasars, redshifts.

\section{Introduction}

The wave patterns of an unknown surface may be deduced from wave patterns in close proximity by using 
a Huygens' construction. In this construction, all points of a wave surface are considered as sources 
synchronous with the wave. The envelope of the \textquotedblleft wavelets\textquotedblright radiated will,
within a short time, create a new wave surface.

A similar construction is obtained replacing the sources of the Huygens' construction by monoatomic or 
polyatomic molecules radiating waves which have the same frequencies and phases as the exciting waves. 
The difference between the Huygens' reconstruction and the original wave form is a function of the finite 
number of sources that introduces discrepancies into the building of a new wave surface. Therefore the 
coherent Huygens' pattern is always perturbed, mixed with incremental amounts of incoherent scattering.

The most common example of coherent scattering is the refraction, whose imperfections produce the 
Rayleigh\footnote{The scattering of light is named here \textquotedblleft Rayleigh 
scattering\textquotedblright if it preserves the frequency, and \textquotedblleft Raman 
scattering\textquotedblright if it changes it.} incoherent scattering (blue of the sky). The amplitude of the 
coherently scattered wave is the sum of the amplitudes of the elementary scattered waves, while the 
incoherently scattered intensity is the sum of the intensities scattered by each molecule. Therefore the 
coherently scattered intensity is N times larger than the incoherently scattered intensity, where N is the 
number of scattering molecules. Since N is usually large, Rayleigh coherent scattering is much more 
dominate than incoherent scattering.

For a long time, only Raman incoherent scattering was observed in the labs, so that the spectroscopists got 
into the habit introducing a stochastic phase correction factor into the off-diagonal elements of the density 
matrix used to study the scattering. The use of lasers has allowed observations of Raman coherent 
scattering, these studies usually involving diffraction of the scattered beam, but in this paper we are 
interested in Raman effects upon wide beams, for which diffraction is negligible.

A method for using Raman scattering to produce a coherent frequency shift without blurring
the image or introducing new lines in the frequency shifted spectra is found in 1968 
\cite{Giordmaine,Treacy}, This technique is developed in \cite{Yan,Weiner,Dougherty,Dhar} and named 
\textquotedblleft Impulsive Stimulated Raman Scattering\textquotedblright (ISRS). Scientists using ISRS 
know that it has no intensity threshold, but it is so difficult to replace the ultrashort laser pulses by the 
pulses which make the usual incoherent light that no one developped \textquotedblleft Coherent Raman 
Effect on time-Incoherent Light\textquotedblright (CREIL). The new name is justified by a qualitative 
difference due to the power of the laser pulses: ISRS is non-linear, the frequency shifts depend on the 
intensity of the laser pulses; on the contrary CREIL does not depend on the intensity of the natural light 
which is much lower than the intensity of the zero point field, except close to very bright stars. 

The properties of CREIL may be deduced from the theory of ISRS, but it is better to develop the 
equations by comparing the basic properties with refraction. This is accomplished in section \ref{CREIL}, 
reproducing and improving already published works \cite{Mor98a,Mor98b,Mor01}.

Section \ref{H2p}, describes the propagation of light in an active gas such as excited molecular hydrogen, 
and explains why this gas cannot be detected even though the quantity of this gas is probably not negligible 
in the intergalactic space.

Section \ref{H} concerns of the propagation of light in atomic hydrogen and some of the conditions 
necessary to produce CREIL. These include excitation by Lyman absorption and the presence of a 
magnetic field.

Hydrogen plasma was chosen because hydrogen is abundant in the Universe, and the long paths between 
the stars and us insure that the weak CREIL (due to the low pressure required to decrease the 
incoherences produced by the collisions) is integrated over a long enough path to create an observable 
redshift. Very simple hypothesis, most of which being not original, are outlined in section \ref{astr} 
proposing an elementary interpretation of many of the spectral features of quasars.

\section{The \textquotedblleft Coherent Raman Effect on Incoherent Light\textquotedblright 
(CREIL).}\label{CREIL}
First we shall examine a coherent light-matter interaction similar to refraction, that is an interaction that 
does not blur the images and the spectra; we will see that it shifts the frequencies, so that its effect may be 
confused with a Doppler shift.
 
\subsection{Condition for no blur of the images: coherent scattering.}
To obtain a strong scattering and avoid blurring the images, the scattered wave surfaces must be identical 
to the wave surfaces of the exciting light beam: The scattering must be coherent.

In Rayleigh coherent scattering, the frequencies and the indices of refraction are the same for the exciting 
and Rayleigh scattered waves, these waves interfere into a single wave. Can we find a similar behavior 
using Raman coherent scattering?

\medskip
If the light is time-coherent, the excitation of a molecule starts with a collision; this is not a problem for a 
Rayleigh scattering because the difference of phase $\phi$ between the exciting and scattered lights 
remains equal to $\pi/2$; however, in Raman scattering $|\phi|$ increases linearly from zero, so that when 
random molecular collisions interrupt the sequence, the light becomes incoherent, producing the effects 
normally observed in Raman scattering. To obtain a coherent scattering, it is necessary to avoid molecular 
collisions during the excitation phase, having, as unique starting point of excitation, the beginning of a light 
pulse: a time-incoherence of the exciting light and a low gas pressure are necessary.
If time-incoherent light is represented by pulses of length $T_{\rm coherence}$, and the collisional free 
time by $T_{\rm collisional}$:
\begin{equation}
T_{\rm collisional}>>T_{\rm coherence}.\label{coher}
\end{equation}

 The mean time $T_{\rm collisional}$ between two collisions in a gas made of identical spherical hard 
molecules is:
\begin{equation}
T_{\rm collisional}=\frac{1}{2Nd^2}\sqrt{ \frac{m}{4\pi kT}} \label{temps}
\end{equation}
 where $N$ is the number of molecules by unit of volume, $d$ their diameter, $m$ their mass and $T$ the 
temperature.

With most gases, $T_{\rm collisional}$ varies little with the molecule: for example, at 300K, for He the 
product $T_{\rm collisional} N$ equals $4.10^{15}s$, while for CO$_2$ it equals 1.86.10$^{15}s$.
In CO2, with a density of the molecules $N = 1.86.10^{23}$, at a pressure 700 Pascals $T_{\rm 
collisional}$ is shorter than $10^{-8}s$. This is only a rough order of magnitude because the molecules 
are not hard particles, so that it is difficult to define a diameter $d$ of molecule for which a collision 
dephases the scattered light.

\subsection{Condition for a strong light-matter interaction: low Raman frequency.}
In refraction, the exciting and scattered beams propagate at the same frequency, thus at the same speed, 
so that they interfere into a single beam. In contrast, in Raman scattering, since the exciting and scattered 
frequencies are different, the indices of refraction $n_e$ and $n_s$ for the exciting and scattered fields are 
also different. Thus, waves diffracted at a distance $x$ on a ray of light have a phaseshift $2\pi(n_e-
n_s)x/\lambda$. A phaseshift equal to $\pi$ occurs for a distance $x$ equal to the \textquotedblleft length 
of coherence\textquotedblright; the amplitudes scattered at this distance cancel by destructive interference; 
this limits the intensity of wide beam Raman coherent scattering.

However, what happens if the Rayleigh scattering produced by transitions inside undivided degenerate 
levels are split into Rayleigh and Raman scatterings by an external field which splits the levels? Examine 
low energy Raman scatterings:

\medskip
Consider first a single Raman Stokes transition.

In the classical theory of the Raman effect, the dipole induced by the incident field is coupled with the 
dipole which radiates the Raman wave; at the beginning of a light pulse, the dipoles are out of phase by 
$\pi/2$, so that the phases of the incident and scattered fields are the same. During the pulse, the phase 
changes because the dipoles have different frequencies.

The interferences patterns of two different frequencies is often observed, for instance between the two 
beams of a Michelson interferometer: when one of the mirrors is moved, there is a Doppler shift in the 
frequency. It can be shown by elementary computation that, if this phaseshift is significantly less than $\pi$, 
the sum of the incident and scattered fields is a single field having an intermediate frequency

The electric field in a pulse of light is the product of a sine function by a slow varying electric field $E(t)$ 
giving the pulse shape; the sine function, for an exciting field of frequency $\nu_e$, may be written $E(t) 
\cos (2\pi \nu_et)$ and, for a field scattered at a frequency $\nu_s$ by a thin layer of thickness $\Delta x$ 
of gas, with the same polarization and the same phase at the beginning of the pulse ($t=0$) :
$E(t) s\Delta x \cos(2\pi \nu _st)$ , where the product $ s\Delta x$ is a small dimensionless coefficient; $ 
s\Delta x$ will be a first order quantity; the sum of the two emerging fields is:
\begin{equation}
D = E(t) (1- s\Delta x)\cos(2\pi \nu _e t)+E(t) s\Delta x\cos(2\pi \nu _st)
\end{equation}
Writing the Raman frequency $ \nu_i =\nu_s - \nu_e$ to eliminate $\nu_s$:
\begin{equation}
\begin{array}{l}
D=E(t)(1- s\Delta x)\cos (2\pi \nu _e t)+\\
+E(t) s\Delta x(\cos (2\pi \nu _et)\cos (2\pi \nu _it)-
\sin (2\pi \nu _et)\sin (2\pi \nu _it))\,.
\end{array}
\end{equation}

Writing the length of the pulse $T_{coherence}$, suppose
\begin{equation}
| \nu_i T_{coherence} |<<1,\,
\end{equation}
that is the Raman period is much larger than the length of the pulses:
\begin{equation}
T_{\rm Raman}>>T_{\rm coherence}.\label{condi}
\end{equation}

We may develop the trigonometric functions of $2\pi\nu_it$ into functions equivalent to them during the 
pulse:
\begin{equation}
\begin{array}{l}
D \approx E(t) \cos (2\pi \nu _et)-2E(t) s\Delta x\pi \nu _it\sin (2\pi \nu _et)\\
+2E(t) s\Delta x(\pi \nu _it)^2\cos(2\pi \nu _et)
+(4/3)E(t) s\Delta x(\pi \nu _it)^3\sin (2\pi \nu _et)\,.
\end{array}
\end{equation}
set:
\begin{equation}
\tan (\psi t)=2 s\Delta x\pi \nu_it;\:(-\pi /2<\psi \leq \pi/2)\,;\label{eqq}
\end{equation}
$
\psi \approx 2 s\Delta x\pi \nu_i \label{phi}
$
is a first order quantity.
\begin{equation}
\begin{array}{l}
D \approx E(t) [\cos (2\pi \nu _et)\cos( \psi t)-\sin (2\pi \nu _et)\sin (\psi t)]
/\cos(\psi t)+\\
+2E(t) s\Delta x(\pi\nu _it)^2\cos(2\pi \nu _et)
+(4/3)E(t) s\Delta x\pi (\nu _it)^3 \sin (2\pi \nu _et)\,.
\end{array}
\end{equation}

In a first order approximation :
\begin{equation}
D \approx E(t) \cos ((2\pi \nu _e +\psi )t).\label{eqs}
\end{equation}
The waves interfere into a single wave within the pulse. Thus, in place of the emergence of a new line 
shifted of $\nu_i $ , the whole incident flux is slightly frequency shifted 
\begin{equation}
\Delta\nu = \psi/2\pi = s\Delta x\nu_i.\label{delnu}
\end{equation}
\subsection{Computation of the ISRS and CREIL redshifts.}\label{num}

An exciting electromagnetic field induces for each Raman transition $i$ a scattered field proportional to an 
element of the tensor of polarizability; averaging this result for all orientations of the gaseous molecules, the 
scattered field is proportional to the trace $\beta_i$ of the tensor of polarizability and has the same 
polarization than the exciting field; therefore, the electric field may be considered as a scalar.

The amplitude ${p_i}$ of the dipole induced in an unit volume for a transition $i$ is proportional to the 
incident electrical field and to the number $N_i$ of molecules per unit of volume in the compatible state: 
$p_i=N_i\beta_i E$. The field scattered at the exciting frequency $\nu_e$ produces the refraction through 
$ P, N$ and $\alpha$ similar to $p_i, N_i$ and $\beta _i$ .

At the thermal equilibrium, $N_i$ is deduced from a Boltzman factor $B_i$, so that the ratio of a Raman 
dipole, with respect to the refracting dipole is:
\begin{equation}
\frac{p_i}{P}=\frac{N_i\beta_i}{N\alpha}=\frac{B_i\beta_i}{\alpha};
\end{equation}
The ratios of scattered amplitudes are the same for a single molecule in any direction, or for a large set of 
identical molecules on an exciting wave surface in the initial direction of propagation\footnote{The relation 
between the coefficients in the two configurations requires the addition of Huygens\textquoteright\hskip 
1mm wavelets by a simple but tedious integration called \textquotedblleft the optical 
theorem\textquotedblright.}.

\begin{equation}
\frac{s_i}{S}=\frac{p_i}{P}=\frac{B_i\beta _i}{\alpha}. \label{rap}
\end{equation}

\medskip
Recall the elementary theory of the refraction index $n$ with our notations:
Set $E\cos(2\pi \nu t)$ the electric field of a wave of frequency $\nu$ as it reaches a thin sheet of gas 
whose thickness $\Delta x$ is a first order small quantity; the absorption through this sheet is neglected so 
that the output field is $E\cos(2\pi \nu (t-\Delta x/c))$; the Rayleigh scattered field is delayed by $\pi /2$, 
the total output field is:
\begin{equation}
E[\cos 2\pi\nu(t-\Delta x/c)+\Delta xS \sin 2\pi\nu (t-\Delta x/c)]
\approx E \cos [2\pi\nu(t-\Delta x/c)-\Delta xS]
\end{equation}
The refraction index $n$ is obtained by an identification of this field with $E \cos 2\pi\nu (t-n\Delta x/c)$, 
giving $n-1=cS/(2\pi\nu)$.

The dynamical dielectric constant $\epsilon$ which equals $1+4\pi N\alpha$ is nearly 1 in a dilute gas, so 
that its square root $n$ equals $1+2\pi N\alpha$; therefore:
\begin{equation}
n-1 = 2\pi N\alpha =cS/(2\pi \nu).\label{ind}
\end{equation}
By equations \ref{rap} and \ref{ind}:
\begin{equation}
\Delta xs_i=\Delta xSB_i\beta _i/\alpha =4\pi^2NB_i\beta_i\nu \Delta x/c\,.
\end{equation}
From equation \ref{delnu}, we get the shift:
\begin{equation}
\Delta \nu=\sum_i[\nu_i\Delta xs_i]=4\pi^2N\sum_i[B_i\beta_i\nu_i]\nu \Delta x/c\,.\label{eqa}
\end{equation}
In a first order development $B_i$ is proportional to $\nu_i$, so that the contribution of transition $i$ 
to the lineshift is proportional to $\nu_i^2$.

\medskip
This formula shows that, neglecting the dispersion of the tensor of polarisability, the relative frequency shift 
$\Delta\nu/\nu$ is constant. But it is very difficult to compute it numerically for the following reasons:

i) it requires the knowledge of a lot of tensors of polarisability;

ii) it is very difficult to apply inequation \ref{condi}.

Therefore we are only able to find a rough order of magnitude of $\Delta \nu/\nu$. To do this, we replace 
the true molecule with a model molecule having a high excitation energy level and two low lying levels of 
energies $E_1$ and $E_2$, close enough to allow a series development of the exponent in the Boltzman 
factor, so that the difference of the populations in the low states is:
\begin{equation}
\begin{array}{l}
N_1-N_2 = N(B_1-B_2) =\\
= N\Bigl[\exp \Bigl (\frac{-E_1}{2kT} \Bigr)- \exp \Bigl(\frac{-E_2}{2kT}\Bigr) \Bigr] \approx 
N\frac{E_2-E_1}{2kT}= N\frac{h\nu_i}{2kT}.\label{Bol}
\end{array}
\end{equation}
 $\beta_i$ requires a transfer of energy from the excited oscillator to the radiating one while $\alpha$ 
does not; but this transfer is generally fast, so that generally $\beta_i$ and $\alpha$ have the same order 
of magnitude; therefore we assume the rough approximation $\alpha = 2\beta_1 = 2\beta_2$. Equation 
\ref{rap} becomes $s_i = B_iS/2 = -SE_i/(4kT)$ for $i =$ 1 or 2. From equations \ref{delnu}, then 
\ref{ind} the frequency shift is
\begin{equation}
\begin{array}{l}
\Delta\nu = \Delta x (s_1-s_2) \frac {E_1-E_2}{h} =\\
= -\frac{\Delta x}{4hkT}( E_1-E_2)^2 \frac{\pi\nu}{c}(n-1)= \frac{\Delta xh\nu_i^2}{2kT} 
\frac{\pi\nu}{c}(n-1). \label{dafin}
\end{array}
\end{equation}

\medskip
In a numerical application:

For all gases $n-1$ is less than an order of magnitude under or over 5.10$^{-4}$ 
in the normal conditions; at a pressure of 400 Pascal which satisfies inequality 
\ref{coher}, $n-1=5.10^{-9}$. To satisfy inequality \ref{condi}, suppose 
$\nu_i=100 MHz$. $\Delta\nu/(\nu \Delta x)=4.10^{-14}m^{-1}$ at 300K. For a 
length of pulse equal to $\Delta t$, Lord Rayleigh's criterion says that the 
frequency difference may be observed if it provides during $\Delta t$ a
difference in phase of $2\pi$ between the original and frequency-shifted pulse. The light beam must 
cross a length of gas X such that $2\pi \Delta\nu\Delta t =2\pi$:
\begin{equation}
\frac{\Delta\nu}{\nu\Delta x}X\nu\Delta t=1 \hskip 3mm{\rm or} \hskip 3mm 
X=\frac{\lambda}{c\Delta t (\Delta\nu/(\nu\Delta x))}\approx 20km
\end{equation}
Our computations of redshift may be optimistic although Rayleigh's criterion is too strong for photoelectric 
measures. Therefore it seems plausible to do an expansive experiment using a long multi-path cell in a 
laboratory.

 \subsection{Properties of the ISRS and CREIL.}
CREIL and ISRS are really a single effect.

In ISRS, the laser energy is so high it overwhelms the collisions that usually de-excitate molecules; in 
CREIL, collisional de-excitation is limited by the low pressure.

In both effects, a radiative de-excitation is necessary, which is provided by a second Raman coherent 
effect involving an other exciting beam. Therefore ISRS and CREIL are not two photons Raman effects, 
which would excite the molecules, but four photons effects, combinations of two simultaneous coherent 
Raman effects. As they do not excite the molecules, these effects are called \textquotedblleft 
parametric\textquotedblright; the molecules play the role of a catalyst in allowing the transfer of energy 
from hot beams of light to cold ones, the temperatures being deduced from Planck's law.

The second Raman effect is provided by a second laser in ISRS, and by the thermal radiation in CREIL. 
In CREIL, this second effect is very strong because all frequencies are very low, so that there are strong 
resonances. Thus the previous evaluation of $\Delta\nu/\nu$ remains valuable.

\medskip
The use of ultrashort, strong laser pulses makes ISRS easily observable: the collisions may be neglected in 
dense matter, the Raman active frequencies may be vibration-rotation molecular frequencies in the infrared. 
In CREIL, the low pressure decreases the probability of scattering; the low Raman frequencies 
correspond to hyperfine transitions which may be:

a) genuine hyperfine transitions involving nuclear spins;

b) transitions between levels split by a Stark or Zeeman effect;

c) transitions in heavy atoms and molecules.

In their low energy states, the light common gases do not have low level state transitions; however, such 
transitions are common in plasma.

\section{Propagation of light in ionised hydrogen.}\label{H2p}
Hydrogen may be ionised by UV radiation into H$_2^+$. This molecule has a complex spectrum due to 
its two nuclear spin and its electronic spin. Homonuclear, it has no permanent dipole, so that, in a first 
approximation, it does not absorb the light by dipolar interactions. A lot of hyperfine structures have been 
observed in microwave band, which show the existence of Raman active transitions in the megaherz range 
\cite{Carrington,Kolos,Leach}: therefore the molecule is active in CREIL; it is stable, but it reacts with 
almost all molecules, so that its persistence requires low pressure which satisfies inequality \ref{coher}. 
Therefore absorption and frequency shift are always simultaneous.

\subsection{Visibility of lines absorbed during a frequency shift}
The absorption of the gas, supposed homogenous, is generally represented by an absorption coefficient 
$k(\nu)$ verifying:
\begin{equation}
{\rm d} \Phi (\nu, x)/ \Phi (\nu, x ) = k(\nu){\rm d} x
\end{equation}

where $\Phi (\nu, x)$ is the flux of energy of a spectral element of frequency $\nu$ in a light beam 
propagating along an $Ox$ axis.

The spectral element observed at a frequency $\nu_X$ has scanned the spectrum from its initial frequency 
$\nu_0$ to $\nu_X$ while it propagated from $x=0$ to $x=X$. Its absorption is:
\begin{equation}
\Delta \Phi =\int_0^X{\Phi(\nu_x,x)k(\nu_x){\rm d} x} \label{absor}
\end{equation}

The sharpest lines get a width larger than the redshift : they cannot be observed individually. As the 
absorption lines of H$_2^+$ are numerous and weak, they cannot be seen: {\it Even though the 
H$_2^+$ is not visible, it can redshift a beam of light}.

\subsection{Detection of CREIL and of its red-shifting gas }\label{reso}
Is it possible to detect that a redshift is produced by CREIL, and to find the nature of the redshifting gas ?

\medskip
A CREIL frequency shift may be written:
\begin{equation}
{\rm d} \nu = \frac{\rho\nu}{q}{\rm d} x
\end{equation}
where $\rho$ is the density of the gas and $q$ a parameter depending on the gas. Assume that the 
composition of the gas is constant; the equation is integrated from the frequencies $\nu_{i0}$ of lines 
emitted or absorbed at $x=0$ to the frequencies $\nu_{iX}$ observed at $x=X$ :
\begin{equation}
\int_0^X\rho{\rm d} x=\int_{\nu_{i0}}^{\nu_{iX}}q\frac{{\rm d} \nu}{\nu} 
=\int_{\nu_{j0}}^{\nu_{jX}}q\frac{{\rm d} \nu}{\nu}=\dots\label{dispersion}
\end{equation}
To explicit that $q$ depends slightly on $\nu$, set $q=q_0+r(\nu)$ where $q_0$ is a constant, and the 
average of the small function $r(\nu)$ is zero. From equation \ref{dispersion}:
\begin{equation}
q_0\bigl[\ln\bigl(\frac{\nu_{i0}}{\nu_{iX}}\bigr)- \ln\bigl(\frac{\nu_{j0}}{\nu_{jX}}\bigr)\bigr]=\\
\int_{\nu_{i0}}^{\nu_{iX}} r(\nu)\frac{{\rm d} \nu}{\nu}-\int_{\nu_{j0}}^{\nu_{jX}} r(\nu)\frac{{\rm 
d} \nu}{\nu}= \int_{\nu_{i0}}^{\nu_{j0}} r(\nu)\frac{{\rm d} \nu}{\nu}-\int_{\nu_{iX}}^{\nu_{jX}} 
r(\nu)\frac{{\rm d} \nu}{\nu}\label{disp2}
\end{equation}
Assume that $\nu_{i0}-\nu_{j0}$ is small; $r(\nu)$ may be replaced by its mean value 
$m(\nu_{i0},\nu_{j0})$ between $\nu_{i0}$ and $\nu_{j0}$, so that the equation becomes:
\begin{equation}
 \ln\bigl(\frac{\nu_{i0}}{\nu_{iX}}\bigr)- \ln\bigl(\frac{\nu_{j0}}{\nu_{jX}}\bigr) =\frac{2(\nu_{j0}-
\nu_{i0}) m(\nu_{i0},\nu_{j0})}{q_0(\nu_{i0}+\nu_{j0})} -\frac{2(\nu_{jX}-\nu_{iX}) 
m(\nu_{iX},\nu_{jX})}{q_0(\nu_{iX}+\nu_{jX})}\label{disp3}
\end{equation}
Assuming that the frequency shift is purely Doppler, the first member of equation \ref{disp3} is zero. 
Otherwise, the simplest explanation for 'Doppler shifting' is CREIL. Observing a spectral 
line emitted by two stars, and assuming that the CREIL is due to the same gas, for a given line the last 
fraction in equation \ref{disp3} has the same value, so that we obtain the variation of $m(\nu_{i0}, 
\nu_{j0})/q_0(\nu_{i0}+\nu_{j0})$ as a function of the star, that is of $\nu_{i0}$ or $\nu_{j0}$; if the 
dispersion of CREIL due to this evaluation is precise enough, the redshifting gas may be characterised.

\section{Propagation of light in atomic hydrogen.}\label{H}
In the excited states of H, the orbital quantum number $\ell$ may be equal or larger than 1, so that, in an 
electric or magnetic field, the energy which depends on the projection quantum number $m$ may produce 
Raman transitions $\Delta m=2$ verifying inequality \ref{condi}. We assume that hydrogen is dissociated 
by sufficient heating (10 000K), and that a Lyman pumping excites the atoms. We also assume that the gas 
is nearly homogenous. In this scenario a variable magnetic field induces a CREIL effect .

\subsection{Lineshape for a relatively high pressure of gas.}\label{BAL}

When the Lyman interactions are strong, an equilibrium is reached between the temperature of the gas and 
the temperature of the light at the resonance frequencies. Therefore, in the absence of a CREIL effect, the 
intensity of the beam at the resonance frequency does not depend on the path in the gas. To find the 
intensity spectrum corresponding to a high magnetic field (and the associated CREIL effect) the magnetic 
field must be evaluated along the path for each of the following conditions:

a) In a high magnetic field: The CREIL redshift rate is high, the temperature of the light at the resonance 
frequency remains constant.

b) As the field strength decreases to zero, as CREIL redshift decreases, the temperature of the light 
approaches the temperature of the gas.

c) While the field strength is zero. the temperature of the light reaches equalibrium with the temperature of 
the gas;a residual redshift widens the line.

d) If the field strength intensifies, the CREIL rate increases.

e) The field strength becomes constant.
 
Since d) and e) are opposite to b) and a) the line has the shape of a trough. It may appear as an emission 
line if the temperature of the gas is higher than the apparent temperature of the source, or else as an 
absorption line in a cool environment. Since an emission line indicates a very high temperature of the light in 
the Lyman frequencies, the emission is highly stimulated and may appear superradiant.

\subsection{Lineshape for a strong but not saturated absorption.}\label{damped}
A Zeeman splitting is usually proportional to a field $H$, so that the CREIL is proportional to the square 
of this field; assuming that the gas is nearly homogenous $x$ is proportional to the mass of gas passed by 
the light beam through a unit of surface. We may write:
\begin{equation}
{\rm d} \nu =A H^2(x){\rm d}x\, ,\label{eqb}
\end{equation}
where $A$ depends on the physical state of the gas. Equation \ref{eqb} may be integrated numerically 
to get $\nu$ as a function of $x$, then equation \ref{absor} is integrated. 

Set $\nu_a$ the absolute frequency of a studied absorption line, $x_n$ a value of $x$ for which $H=0$ 
and $X=x-x_n$; mark a spectral element of the light by its frequency $f$ for $X=0$. Suppose now that 
the absorption is low and that the linewidth is purely Doppler, so that without a field we would have the 
absorption at a frequency $\nu$:
\begin{equation}
{\rm d}\Phi (\nu)=-\exp(-a(\nu-\nu_a))^2{\rm d}x\label{13} 
\end{equation}
equation \ref{eqb} , written with a convenient coefficient $b$ becomes, for $X$ small:
\begin{equation}
{\rm d}\nu/{\rm d}X=bX^2.
\end{equation}
Integrating
\begin{equation}
\nu=bX^3/3+f.\label{15}
\end{equation}
From equations \ref{13} and \ref{15} the variation if the intensity of the spectral element is:
\begin{equation}
{\rm d}\Phi (f)=-\exp(-a(bX^3/3+f-\nu_a)^2) {\rm d}X
\end{equation}

\begin{figure}
\begin{center}
\includegraphics[height=5 cm]{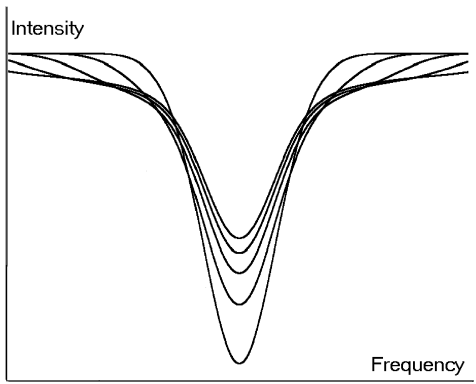}
\end{center}
\caption{Computed shape of absorption lines.}
\label{forme}
\end{figure}

\medskip
Figure \ref{forme} shows lineshapes computed with $a$ constant, a zero and four non-zero values of $b$; 
the half intensity width is nearly constant while the absorption remains large outside of this region in the feet 
of the line: the line is damped. Taking into account only the fast changing intensities in the spectrum, that is 
neglecting the base of the lines, the pseudo lines appear as sharp as a line stretched by the thermal Doppler 
effect.

\subsection{Propagation with low pressure and low light intensity.}\label{forest}
Assuming that the gas and the magnetic field are homogeneous and that the intensity of the absorption is 
constant, name \textquotedblleft critical length\textquotedblright $L_c$ the length of path required to 
produce a redshift equal to the mean width $\delta\nu$ of the lines whose absorption is necessary to get 
the redshift. At both ends of a path long of $L_c$, the intensity has a same value $I_c$ so that our 
hypothesis are self-coherent.

Assume that the intensity is decreased, so that the redshift is decreased; after a path of the previously 
defined length $L_c$, the spectral elements which pump the gas are not fully renewed, so that the redshift 
is lowered. Iterating, the redshift disappears unless it exists atoms the redshifting power of which does not 
require a pumping; as this possible residual redshift has a low intensity, the absorption at {\it all} eigen 
frequencies of all components of the gas are large. Therefore, a previous absorption may start an extinction 
of the redshift and the writing of many absorption lines.

On the contrary, assume that the intensity is equal to $I_c$; a redshift is proportional to the surface number 
(number per unit of surface) of excited atoms, so that the surface number $N_e$ of excited atoms along 
$L_c$ is well defined. An increase of the intensity is an increase of $I_c$ because the density of excited 
atoms increases, so that $L_c$ decreases; assuming a linearity, the absorption of a line has a value 
deduced from $N_e$, which does not depend on the incident intensity. These absorptions move the base 
line of the spectrum, so that the contrasts of the weak lines already written in the spectrum are increased.

\medskip
Consequently, a complicated interaction between spectra written at various redshifts appears: The 
\textquotedblleft Lyman forest\textquotedblright may be a chaotic spectroscopic effect depending strongly 
on an initial setting of a spectrum, then on unpredictable slow variations of the properties of the halo.

\section{Applications to astrophysics.}\label{astr}
The Universe is made up of mostly hydrogen; in the usual interpretation of the spectra of quasars, the 
intergalactic space contains areas of hot atomic hydrogen, but in our following interpretation this hot 
hydrogen is in the extended atmosphere of quasars. We therefore assume that the intergalactic space is 
cold and contains molecular hydrogen partly ionised by UV radiation from the stars. 

\subsection{The cosmological redshifts.}
Suppose that H$_2^+$ has the same efficiency in CREIL as the gas considered in \ref{num}. What 
density $N$ would be required to provide the Hubble redshift without a Doppler, gravitational or 
expansion contribution?
\medskip
For a moderate redshift, Hubble's law is
\begin{equation}
\frac{\Delta\nu}{\nu L} = \frac{1}{L}\sqrt{\frac{c-LH_0}{c+LH_0}}-\frac{1}{L} \approx –
\frac{H_0}{c}\label{hubble}
\end{equation}
Using the value H$_0/c=10^{-26}$, we obtain a pressure of $10^{-14}$Pa, that is $N\approx3.10^6$ 
molecules per cubic metre. If the temperature of the gas is 3K in place of 300K, it remains $N=3.10^4$ 
molecule m$^{-3}$ that is 30 molecules per litre.

\subsection{A model of quasars and Seyfert galaxies.}
The Lyman forest of the quasars which demonstrates the existence of hot atomic hydrogen is not readily 
observed in low redshifts quasars. However some of the spectral traits are observed and this may indicate 
there is intrinsic redshift in these smaller quasars as well, indicating they may be much closer than expected. 
If this is true, these quasars may be very close \cite{Petitjean,Shull}. Consequently their central engine 
does not radiate an enormous energy, it may be a small star (a neutron star?) heated by the fall of an 
accretion disk slowed by a relatively dense halo of atomic hydrogen. We propose that the density of the 
halo, its metallicity, and, with possible exceptions, its temperature decrease with the distance to the star. 
The slow change of properties of the gas is observed, but hard to explain for the clouds of the regular 
model \cite{Tytler}.

\medskip
Observing quasars, Webb et al. \cite{Webb} measured that the relative frequency shift of lines absorbed 
by the same multiplet, thus absorbed by the same atoms are not equal. They wrote that it is due to a 
variation of the fine structure constant. A CREIL in gases between the quasars and us is a simpler 
explanation. As all frequencies are known, equation \ref{disp3} gives a relation between $ 
m(\nu_i,\nu_j)/q_0$ and $ m(\nu_{io},\nu_{jo})/q_0$. Thus, as indicated in subsection \ref{reso}, 
numerous, careful spectroscopic measures could measure the resonances of the CREIL, and possibly 
detect their origin which, in this case, is very likely atomic hydrogen.
\subsection{Correlation between the broad lines and the radio quietness of the quasars (and 
Seyfert galaxies).}
\begin{figure}
\begin{center}
\includegraphics[height=5 cm]{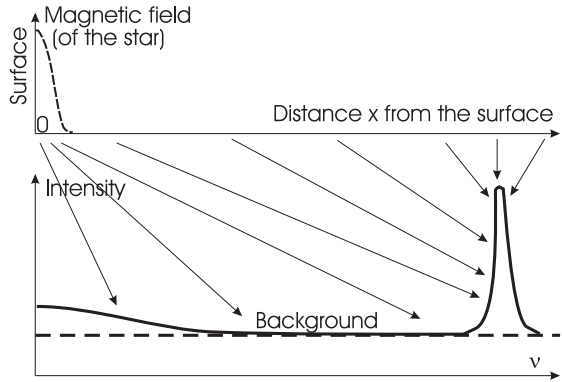}
\end{center}
\caption{The high redshift emission lines of the quasars. The arrows show the correspondence between the 
distance to the star x and the frequency. In the absence of magnetic field (and therefore no CREIL),, the 
emission along a long path concentrates at a frequency, producing a strong emission line. }
\label{surface}
\end{figure}
Many authors think that the quasars and Seyfert galaxies have an accretion disk; set $\theta 
(0\leq\theta\leq\pi/2)$ the angle between the axis of the disk and the line of sight to the Earth. The physics 
of the disks is very complex \cite{Beloborodov}; the star and the disk can produce X rays whose 
absorption by the disk \cite{Brandt,Reeves,Mathur} is observed for $\theta\approx\pi/2$.

The radio emission of the quasars is generally attributed to the interaction of jets with magnetic fields; if this 
emission is produced by electric discharges in or at the surface of the disk, its intensity decreases with 
$\theta$ down to nearly zero for $\theta=\pi/2$ because these discharges are flat.

As the star is not a blackbody, the hot gases in the lower atmosphere radiate emission lines. Very close to 
the quasar, the magnetic field produces CREIL; Here the effect is very strong producing a rapid redshift, 
but the spread emission line is nearly invisible (figure \ref{surface}). Where the magnetic field disappears, 
the strong emission lines traditionally used to define the redshift of the quasar are written sharply into the 
spectrum.

\begin{figure}
\begin{center}
\includegraphics[height=5 cm]{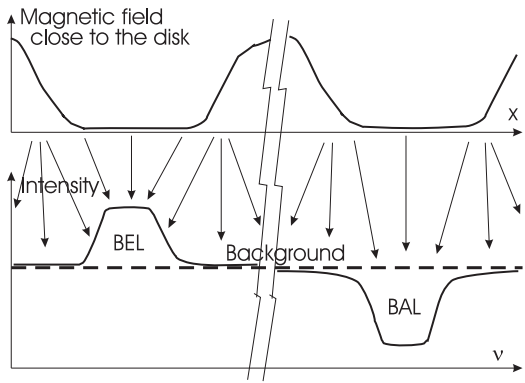}
\end{center}
\caption{Shapes of the broad lines.The emission or absorption reaches the 
equilibrium between the temperature of the gas and the temperature of the light 
at the resonance frequency.}
\label{figbal}
\end{figure}

If $\theta\approx\pi/2$, the light propagates close to the disk which produces a variable magnetic field 
\cite{Welter,Kool}. Here, the conditions exist described in subsection \ref{BAL}, trough are written into 
the spectrum, first in emission, then in absorption (Figure \ref{figbal}).
The presence of broad lines, as observed \cite{Stocke}, is not compatible with detection of strong radio 
emissions even though all other spectral properties are similar \cite{Weymann}.

In radio-loud conditions ($\theta\ll\pi/2$), there is no magnetic field and no CREIL, the emission, then 
absorption which corresponds to the broad lines are confused as an excess of absorption near the 
emission lines used to define the redshift of the quasar \cite{Briggs,Anderson}. In this case, the 
propagation does not introduces a redshift, while it does in the other case; consequently the thermal 
radiation near the core is less amplified. This is why the the \textquotedblleft dust 
emission\textquotedblright is lower than in BAL quasars \cite{Omont2}.

\subsection{The damped absorption lines and the Lyman forest.}
\begin{figure}
\begin{center}
\includegraphics[height=5 cm]{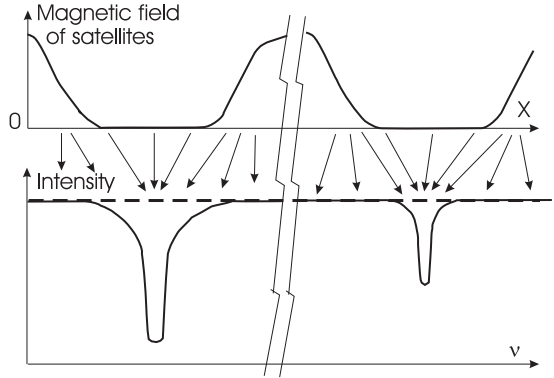}
\end{center}
\caption{Damped and forest lines. Figure \ref{forme} provides better shapes.}
\label{fforest}
\end{figure}
Many Lyman absorption lines are observed in the spectra of the quasars; in the standard model, these lines 
are absorbed by clouds of hydrogen in the interstellar medium. But it is 
difficult to explain the confinement of these clouds \cite{Shull}.
The Seyfert galaxies get their name from a proposed existence of many satellites. These satellites may be 
surrounded by magnetic fields, so that the line of sight crosses a variable magnetic field. As described in 
subsection \ref{damped}, damped lines are \textquotedblleft written in\textquotedblright with a 
lower redshift than the broad lines(Figure \ref{fforest}).

\medskip
It seems difficult to have a density of satellites large enough to explain the large number of lines in the 
Lyman forest \footnote{However the short decreases of intensity of the quasars may be due to 
occultations of the star by satellites.}. A linking of absorption patterns written when a shifted, already 
written line reaches a Lyman line (subsection \ref{forest}) may produce the quantification observed by 
Burbidge and Hewitt \cite{Burbidge,Hewitt}, Bell and Comeau \cite{Bell,Comeau}: a coincidence of the 
Ly$_\alpha$ line with shifted Ly$_\beta$ and Ly$_\gamma$ lines corresponds to $z=0.185$ and 
$z=0.251$ respectively, and these values are the products by 3 and 4 of the fundamental redshift 0.062 
observed experimentally by these authors.

\subsection{The dust.}
Where the CREIL effect is greatest, the absorption is weak, but the sum of the absorptions by all lines is 
not negligible; it may be confused with absorption by dust. The energy lost by the redshifts heats the 
thermal radiation, just like hot dust. This solves a paradox, that the bright, much redshifted objects appear 
dusty, while at the same time the dust is not burnt in the plasma, or rejected by the pressure of radiation 
\cite{Shull95,Omont,Priddey}.

\section{Conclusion}
It is the consensus among astrophysicists and cosmologists that all observed redshifts are the result of 
Doppler, expansion or gravitational effects. Part of the justification is that Coherent Raman Scattering of 
time-Incoherent Light (CREIL) has not been demonstrated in the laboratory, and to do so would require 
an expansive experiment. However commonly used Impulsive Stimulated Raman Scattering (ISRS) differs 
qualitatively from CREIL only by the nonlinearity caused by the power of the pulsed lasers used to 
produce this effect.

The near vacuum of space, and the high flux expanses near quasars provides the right conditions for 
CREIL. CREIL provides a very simple explanation to many observations which defy expansion or 
gravitational solutions, even with the introduction of strange concepts such as dark matter.

\medskip
This work is only an outline, both the physics and the astrophysics must be explored: For instance, the 
residual redshift in the Solar spectra, after correction of the local Doppler and gravitational effects, are 
proportional to paths in atomic, magnetised, and therefore CREIL active hydrogen.

\section{Acknowledgement}
I thank Jerry Jensen very much for many corrections of the manuscript he 
suggested.

\end{document}